\documentclass[conference]{IEEEtran}
\usepackage{lettrine}
\usepackage{graphicx} 
\usepackage{textcomp}
\usepackage{amsmath}
\usepackage{amssymb}
\usepackage{gensymb}
\usepackage{subcaption}
\usepackage{circuitikz}
\usepackage{tikz}
\usepackage{bm}
\usepackage{float}
\usepackage{hyperref}
\usepackage[acronym, toc]{glossaries} 
\usetikzlibrary{shapes.geometric, arrows, positioning}
\title{Design and Analysis of a 2D Vernier Structure}

\author{
    \IEEEauthorblockN{Husna Yildiz, Rick Helmer}
    \IEEEauthorblockA{
        TU Delft, Email: rickhelmer@tudelft.nl;\\
        University, Email: husnayildiz@tudelft.nl
    }
}

\begin{document}
\pagenumbering{arabic}
\thispagestyle{plain}    
\pagestyle{plain}  

\maketitle

\begin{abstract}
This paper presents the design and simulation of a 2D Vernier Time-to-Digital Converter (TDC) targeting less than 20 picosecond time resolution and 5-bit output resolution. The TDC comprises voltage-controlled delay buffers, NAND-based SR latches with reset functionality, and a calibration loop using a phase-frequency detector (PFD) and charge pump. The calibration loop dynamically adjusts delay mismatches to maintain consistent timing across process-voltage-temperature (PVT) corners, significantly improving linearity. Simulations show that the TDC achieves a resolution of 16.9 ps for tt corner and maintains DNL and INL below 0.5 LSB under all tested PVT conditions.
\end{abstract}

\section{Introduction}
Time-to-digital converters (TDCs) play a crucial role in digital phase-locked loops (PLLs), high-resolution timing systems, and various precision measurement applications. Among various architectures, the two-dimensional (2D) Vernier TDC offers an approach by quantizing time differences through delay mismatches between two orthogonal delay lines arranged in a grid, known as the Vernier plane.

In the 2D Vernier method, time intervals are quantized by comparing signal propagation through delay lines X and Y. When each stage delay is an integer multiple of the desired resolution, the quantization step $\Delta t_{\text{res}}$ is determined by the greatest common divisor (GCD) of the individual stage delays.

Compared to traditional linear TDCs, the 2D Vernier structure significantly shortens delay lines, reducing jitter and power consumption while improving linearity. Reusing delay taps across both dimensions also enhances quantization density and range. This makes the 2D architecture well-suited for high-resolution systems, offering an improved tradeoff between power, area, and resolution.

However, the approach is not without challenges. Finite delay lines can result in non-uniform quantization, making precise delay matching and calibration critical. Achieving low DNL and INL often requires a larger Vernier plane to minimize abrupt diagonal transitions in the output codes \cite{2dvernierplane}.

\subsection{Design Objectives}
The goal of this project is to design a 2D Vernier time-to-digital converter that meets the following requirements strictly:
\begin{itemize}
    \item Time resolution better than 20 picoseconds
    \item 5-bit resolution (32 distinct output codes)
\end{itemize}
In addition to the core requirements, the design aims to meet the following performance targets:
\begin{itemize}
    \item DNL and INL below 0.5 LSB for high linearity
    \item Optimized Figure of Merit (FoM), defined as:
    \[
        \text{FoM} = \frac{\text{energy\_per\_conversion}}{2^{\text{ENOB}}}
    \]
\end{itemize}
Although area is not explicitly part of the FoM used, minimizing area remained a core design goal due to its strong impact on power. Therefore, area efficiency was considered throughout the design process, even if not captured in the final FoM formula.

\subsection{Project Boundaries}
The scope of this project is limited to the design and simulation of a stand-alone 2D Vernier TDC. The following boundaries apply:

\subsection*{Included}
\begin{itemize}
    \item Design and implementation of the 2D Vernier TDC
    \item Optional delay calibration loop using a charge pump and delay detector
    \item Simulation and analysis under various PVT corners and mismatch conditions using Spectre
    \item Evaluation of key metrics: time resolution, energy per conversion, DNL/INL, area (sum of transistor widths), and FoM
\end{itemize}

\subsection*{Excluded}
\begin{itemize}
    \item Layout and post-layout simulations
    \item Full system-level integration; the TDC is treated as a stand-alone block
    \item Clocking design
\end{itemize}

\subsection{Optional Enhancements and Design Boundaries}
Although not required, a calibration loop was implemented to compensate for delay mismatches between the Start and Stop paths across PVT variations. This loop, consisting of a delay detector and charge pump, enhances accuracy and robustness.
Design boundaries introduced by this enhancement include:
\begin{itemize}
    \item The calibration loop must remain stable and converge reliably across all corners
    \item Slight increase in power consumption and circuit complexity
\end{itemize}
Despite these trade-offs, the calibration loop ensures accurate signal alignment on the Vernier plane, improving linearity and overall system reliability.

\subsection{Report Structure}
This report presents the design of a 2D Vernier TDC with a resolution better than 20 ps and 5-bit output. The design is optimized for high performance in terms of INL, DNL, area, power consumption, and figure of merit. Section II explains the design methodology, details the sub-blocks that make up the 2D Vernier structure. Section III presents the simulation results. Section IV discusses potential directions for future work. Finally, section V concludes with a summary of key findings.

\section{Design Methodology}

The presented design implements a two-dimensional Vernier Time-to-Digital Converter, which compares signal propagation times across two independent delay lines, referred to as line X and line Y. The structure forms a 2D grid where each intersection represents the differential delay between a tap from line X and a tap from line Y. In this design, 9 delay stages are used in line X , 5 delay stages in line Y and 2 delay stages for balanced loading, resulting in 45 distinct intersection points. Each of these intersections is connected to an SR latch functioning as a time comparator.

The architecture is implemented in a 180\,nm CMOS technology. It consists of several key building blocks: voltage-controlled delay buffers, SR latches and a calibration loop composed of a phase-frequency detector (PFD) and a charge pump. The voltage-controlled buffers are used in line X and Y to enable fine control of the delay, and the calibration system ensures that timing alignment is maintained between the two lines.

The resolution of the TDC is defined by the delay difference between the X and Y delay lines. In this implementation, the delay of a single stage in line Y is denoted by \( \tau \), while each stage in line X is deliberately made slower. The ratio between the stage delays is set to \( \tau_X / \tau_Y = 5/4 \), meaning that the delay per stage in the X line is \( \tau_X = \frac{5}{4} \tau \). This configuration ensures that the accumulated delay of 5 stages in line Y equals the delay of 4 stages in line X. Therefore, the calibration point occurs at position X = 4 and Y = 5, where the total delays are equal and the differential delay is zero.

The resolution \( \Delta \) is defined as the difference between the delays of a single X and Y stage:
\[
\Delta = \tau_X - \tau_Y = \frac{5}{4} \tau - \tau = \frac{1}{4} \tau.
\]
This implies that this architecture of the TDC can resolve time differences down to one-quarter of the Y-stage delay.

The calibration loop actively adjusts the control voltage applied to the voltage-controlled delay buffers in the X line to maintain this delay relationship, ensuring that the resolution remains constant and accurate over process and temperature variations.

The full-scale range of the TDC spans from a relative delay index of -3 to 34, providing 38 quantization levels in total. However, only 32 of these levels are used to achieve the 5-bit resolution required by the design. The remaining 7 SR latches serve as dummy comparators, ensuring uniform loading across all taps in the delay grid.
\begin{figure}[H]
    \centering
    \includegraphics[width=\linewidth]{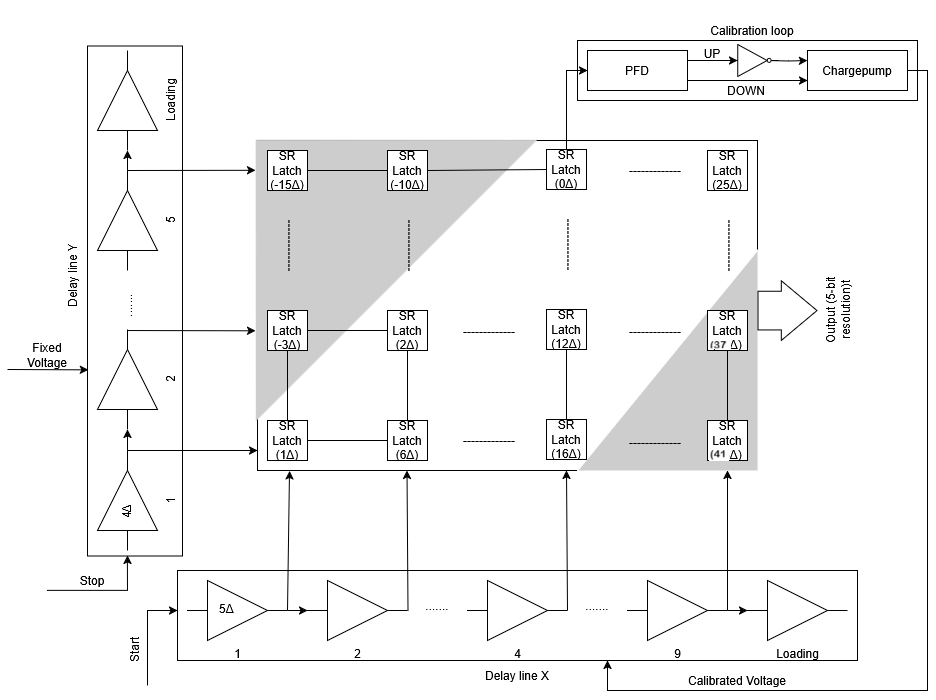}
    \caption{Architecture of the proposed 2D Vernier TDC. The SR latches in the gray zone are included for loading purposes.}
    \label{fig:vernier_architecture}
\end{figure}

\subsection{Voltage-Controlled Delay Buffer}
In the Start and Stop signal paths, current-starved buffers are made to achieve delay control. Each buffer comprises two inverters, with focusing on minimizing propagation delay while preserving signal integrity. Through empirical tuning, optimal sizing was determined to achieve the fastest propagation.

To have more control on the delay, a current-starving NMOS transistor is added. An initial approach using separate starving transistors for both inverters led to signal degradation. Therefore, the final design utilizes a single current-starving transistor.

The bias voltage for the X-line buffers are set by a calibration loop, which ensures that the Start and Stop paths consistently maintain the desired 4:5 delay ratio across all process, voltage, and temperature (PVT) corners. 

Table~\ref{tab:buffer_sizing} summarizes the transistor sizes used for the buffers. All buffers in the delay lines were designed with identical sizing. To ensure consistent delay behavior, dummy loads were added to the inputs of the first and last buffers in both the Start and Stop delay lines, ensuring that all buffers experience identical loading conditions and thus exhibit uniform delay.

\begin{figure}[H]
\centering
\includegraphics[width=0.8\linewidth]{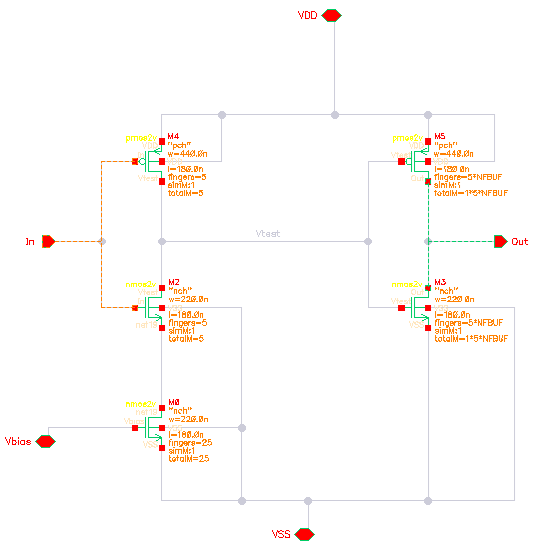} 
\caption{Schematic of the voltage-controlled delay buffer used in the Start and Stop paths.}
\label{fig:buffer_schematic}
\end{figure}

\begin{table}[H]
\centering
\caption{Transistor Sizing for Buffer and Current Starving Elements}
\label{tab:buffer_sizing}
\begin{tabular}{|l|c|c|c|c|}
\hline
\textbf{Component} & \textbf{Type} & \textbf{W (nm)} & \textbf{L (nm)} & \textbf{Fingers} \\
\hline
1\textsuperscript{st} inverter PMOS         & PMOS & 440 & 180 & 5  \\
1\textsuperscript{st} inverter NMOS         & NMOS & 220 & 180 & 5  \\
2\textsuperscript{nd} inverter PMOS         & PMOS & 440 & 180 & 15 \\
2\textsuperscript{nd} inverter NMOS         & NMOS & 220 & 180 & 15 \\
Current starving NMOS                       & NMOS & 220 & 180 & 25 \\
\hline
\end{tabular}
\end{table}
Figure~\ref{fig:buffer_schematic} presents the schematic of the voltage-controlled delay buffer used in both X and Y delay lines.

\begin{figure}[H]
    \centering
    \includegraphics[width=1\linewidth]{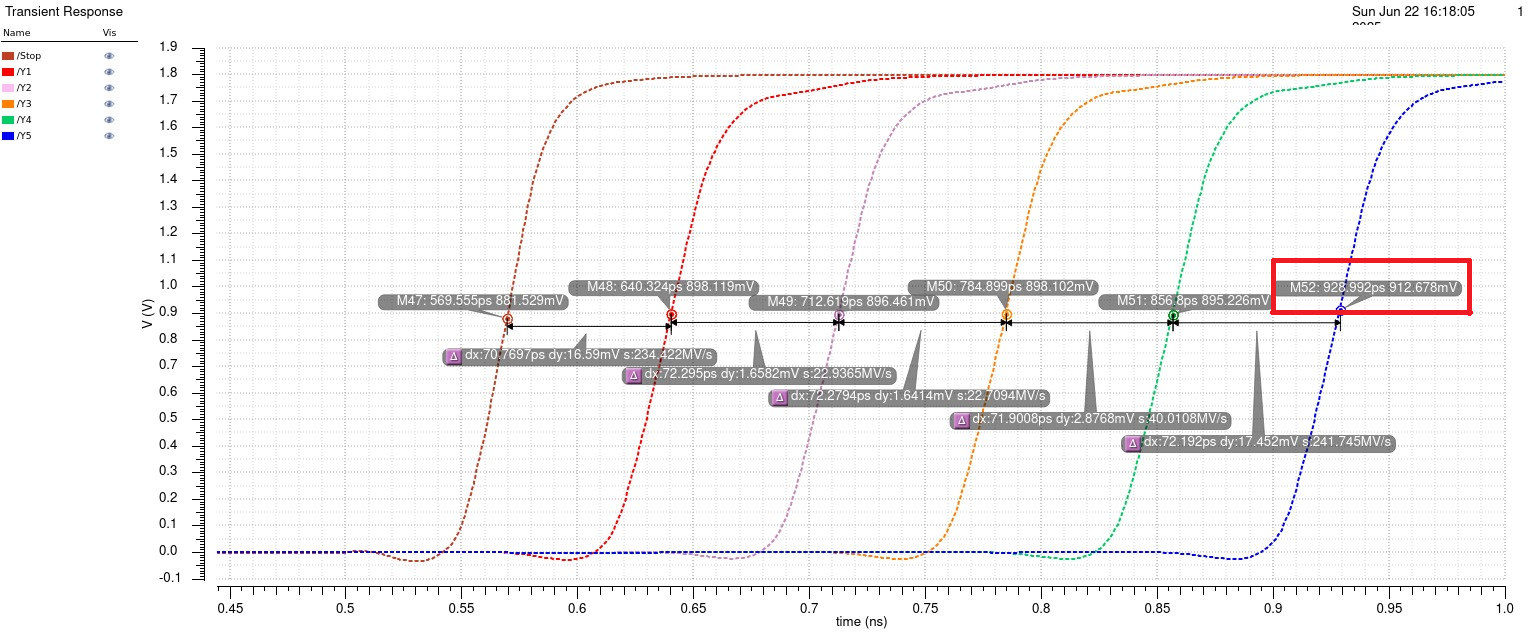}
    \caption{Simulation illustrating the propagation of the stop signal along the y line.}
    \label{fig:propagationStop}
\end{figure}

\begin{figure}[H]
    \centering
    \includegraphics[width=1\linewidth]{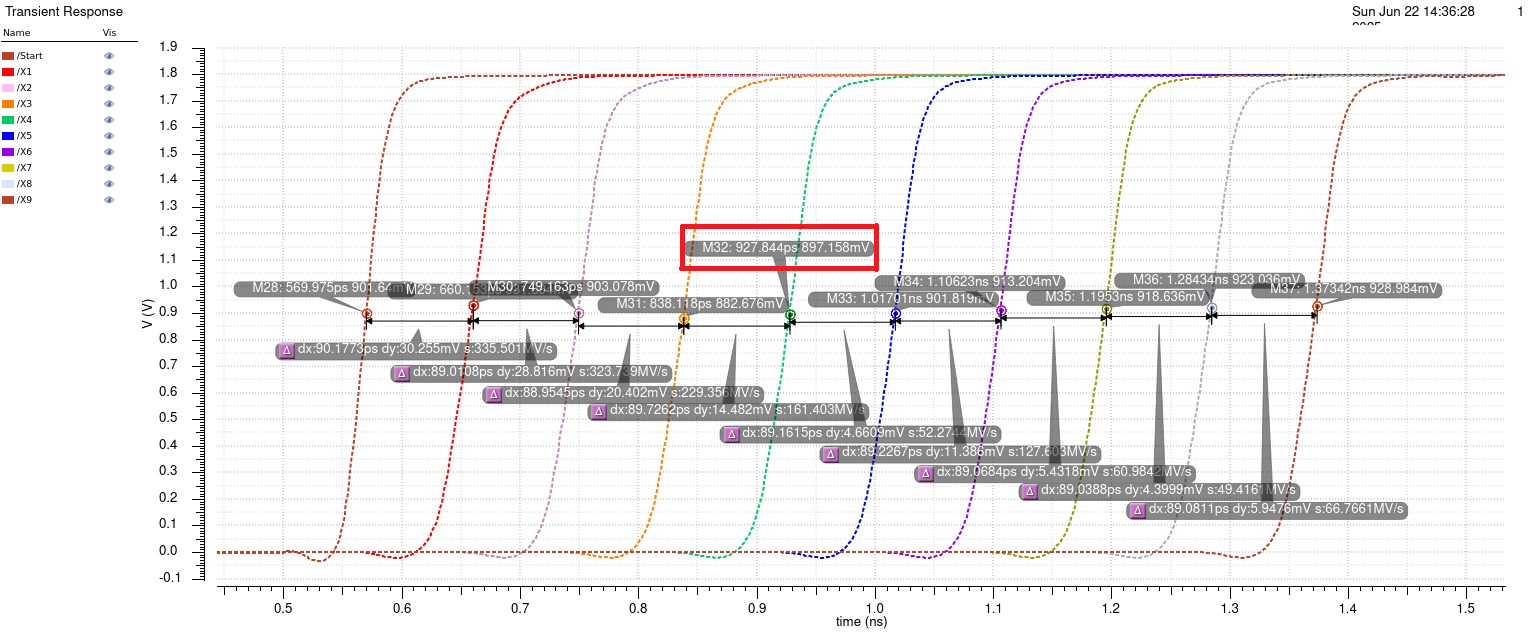}
    \caption{Simulation illustrating the propagation of the start signal along the X line.}
    \label{fig:propagationStart}
\end{figure}

Figures \ref{fig:propagationStop} and \ref{fig:propagationStart} present the simulation results. Initially, the Stop signal propagation along the Y line is depicted in Figure \ref{fig:propagationStop}. The simulation shows an average delay of 71 ps (=\(\tau_Y\))between successive buffer stages. Similarly, the Start signal propagation, shown in Figure \ref{fig:propagationStart}, exhibits an average delay of 89 ps (=\(\tau_X\)) per buffer stage.

A key observation is the effectiveness of the calibration loop, which dynamically adjusts the delay elements to maintain the target \( \tau_X / \tau_Y = 5/4 \). This is confirmed by the close agreement: $89 ps \cdot 0.8 \approx 71.2 ps$, closely matching the measured Stop delay.

Finally, the point at which the Start and Stop signals simultaneously arrive at position X = 4 and Y = 5 in red in Figures \ref{fig:propagationStart} and \ref{fig:propagationStop}, validates the correct operation of the calibration loop.


\subsection{Time Comparator: SR Latch}
\begin{figure}[H]
    \centering    \includegraphics[width=0.9\linewidth]{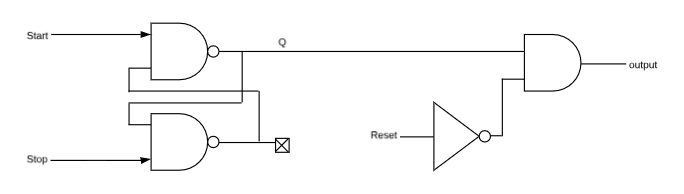}
    \caption{Schematic of implemented SR latch with reset stage.}
    \label{fig:srLatch}
\end{figure}

\begin{figure}[h!]
    \centering
    \includegraphics[width=1\linewidth]{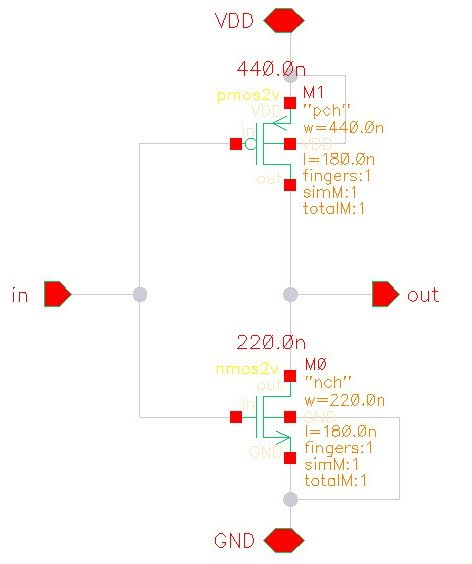}
    \caption{Schematic of Inverter}
    \label{fig:inverterSchematic}
\end{figure}

\begin{figure}[h!]
    \centering
    \includegraphics[width=1\linewidth]{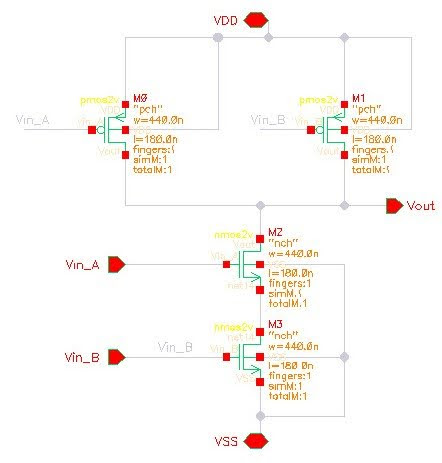}
    \caption{Schematic of NAND gate}
    \label{fig:nandgate}
\end{figure}

\begin{figure}[h!]
    \centering
    \includegraphics[width=1\linewidth]{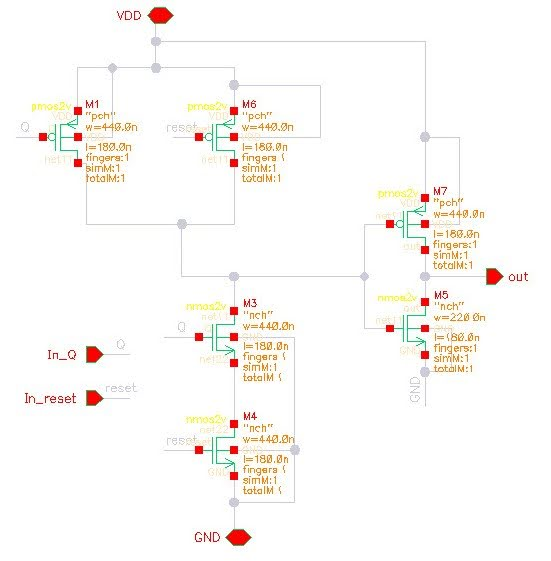}
    \caption{Schematic of AND gate}
    \label{fig:andgate}
\end{figure}

Figure \ref{fig:srLatch} shows the SR latch used in our 2D Vernier TDC design. Figures \ref{fig:inverterSchematic} to \ref{fig:andgate} show the schematics of the NAND, AND, and inverter gates. It’s a NAND gate-based SR latch, where Start and Stop are the inputs.

Each SR latch functions as a time comparator and detects the relative arrival time of the signals along the X and Y delay lines. Specifically:
\begin{itemize}
    \item If the Start signal (from the X-line) arrives before the Stop signal (from the Y-line), the SR latch is set and its output goes high.
    \item If the Stop signal arrives earlier, the latch is output goes low.
\end{itemize}
The outputs from these latches collectively encode the TDC result. In total, 32 SR latches are used to represent the valid 5-bit output range from $\Delta = 0$ to $\Delta = 31$. Each latch corresponds to a quantization point. For example, if the first five latches corresponding to $\Delta = 0$ to $\Delta = 4$ are high and the rest are low, it indicates that the Stop signal is 4 delay steps later than the Start signal. In other words, the time difference between the two signals is $4\Delta$.

To ensure robust operation and reusability across multiple measurements, a reset function is included in each latch. The right side of the circuit includes an additional logic stage that gates the output $Q$ using an external Reset signal. When Reset is low, this stage is transparent and the output reflects the latch state. When Reset is high, the output is forced to zero, clearing all latch states between measurements and preventing residual outputs.

\begin{figure}[H]
    \centering
    \includegraphics[width=1\linewidth]{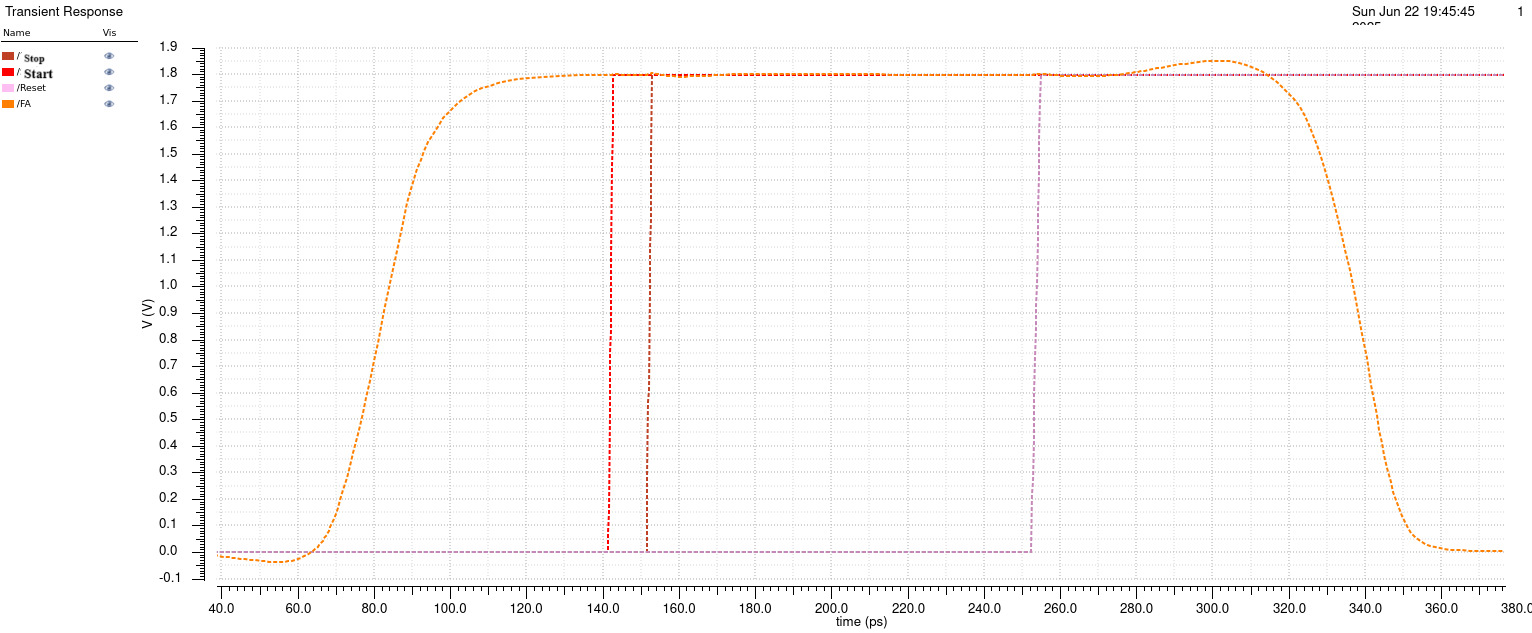}
    \caption{Simulation of the SR latch, where the rising edge of the start signal sets the output high, and the reset being high results the 
output to low.}    \label{fig:simulationResult}
\end{figure}

Transistor sizes used in the latch and logic stages are summarized in Table~\ref{tab:sizing}. Simulation results shown in Figure~\ref{fig:simulationResult} confirm the correct behavior: the Start signal rising sets the latch output high, and the Reset signal (shown in pink) pulls the output back to zero as expected.

\begin{table}[H]
\centering
\caption{Transistor Sizing Summary}
\label{tab:sizing}
\begin{tabular}{|l|l|c|c|c|}
\hline
\textbf{Component} & \textbf{Type} & \textbf{W} & \textbf{L} & \textbf{Fingers} \\
                  &               & \textbf{[nm]} & \textbf{[nm]} & \\
\hline
Pull-up PMOS NAND      & PMOS  & 440 & 180 & 1 \\
Pull-down NMOS NAND    & NMOS  & 440 & 180 & 1 \\
Pull-up PMOS inverter  & PMOS  & 440 & 180 & 1 \\
Pull-down NMOS inverter & NMOS  & 220 & 180 & 1 \\
AND gate PMOS          & PMOS  & 440 & 180 & 1 \\
AND gate 1st stage NMOS & NMOS  & 440 & 180 & 1 \\
AND gate 2nd stage NMOS & NMOS  & 220 & 180 & 1 \\
\hline
\end{tabular}
\end{table}

\subsection{Calibration}
The linearity  of the 2D Vernier TDC critically depends on precise timing differences between the delay elements of the two delay lines. To achieve and maintain this precision, a calibration loop was implemented. This loop primarily comprises a phase-frequency detector (PFD) and a charge pump. It continuously monitors the timing difference between signals propagating through delay lines X and Y, adjusting the delay of the X line via voltage-controlled buffers to ideally maintain a zero timing difference at the calibration point X = 4 and Y = 5. During calibration, the Start and Stop signals are launched simultaneously without any intentional delay between them. This allows the calibration loop to correctly interpret any measured delay difference as being due to mismatches between the delay elements, rather than external skew.

\subsubsection{Phase-Frequency Detector}
A phase-frequency detector with zero dead zone is essential for accurate calibration, as the targeted timing differences are on the order of picoseconds. The selected PFD architecture \cite{calibration_loop} employs only eight transistors and provides completely dead-zone-free operation. Due to the absence of a dead zone, the PFD reliably detects minimal phase differences between the input signals.

In operation, the chosen PFD immediately outputs a high-level signal (UP or DN), depending on the 2 input signals, thus directing the charge pump accordingly. Specifically, the UP signal becomes high when the signal from the X line arrives later than the signal from the Y line. Conversely, the DN signal becomes high if the X line signal arrives earlier. This allows the calibration loop to adjust delays and align both signals accurately.

Its simplified transistor configuration significantly reduces power consumption and enhances operating frequency, making it ideal for high-speed, low-power calibration applications \cite{calibration_loop}.

\begin{figure}[h!]
    \centering
    \includegraphics[width=0.45\textwidth]{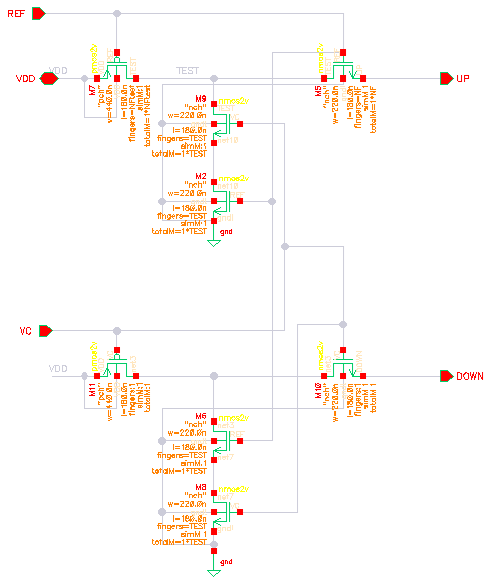}
    \caption{Schematic of the implemented Phase-Frequency Detector (PFD).}
    \label{fig:PFD_schematic}
\end{figure}

Figure \ref{fig:PFD_schematic} shows the schematic of the implemented PFD.

\subsubsection{Charge Pump}
The charge pump converts UP and DN signals from the PFD into a control voltage, used to tune the voltage-controlled delay buffers in line X. Accurate calibration requires the charge pump to have matched sourcing and sinking currents and minimal output ripple.

The implemented charge pump uses a conventional topology \cite{chargepumpreview} with a PMOS transistor for charging and an NMOS transistor for discharging, both controlled by UP and DN signals, respectively. An inverter ensures proper logic levels for the PMOS transistor. Transistor sizing and simulations verified balanced charging and discharging currents.

A standard current mirror configuration provides a constant bias current of 1 mA, mirrored to the charge and discharge branches, minimizing mismatch.

The loop filter consists of a parallel RC network: the resistor and first capacitor integrate the charge pump current, while the second capacitor stabilizes the control voltage and reduces ripple. Although some delay mismatches were observed due to asymmetric paths, transistor sizing was carefully optimized to minimize this effect.

\begin{figure}[H]
    \centering
    \includegraphics[width=0.5\textwidth]{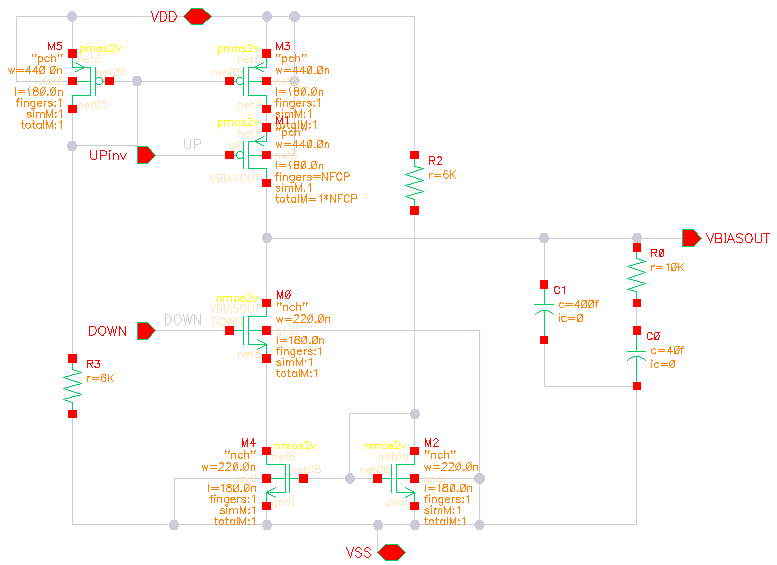}
    \caption{Schematic of the implemented charge pump.}
    \label{fig:ChargePump_schematic}
\end{figure}

The schematic of the implemented charge pump is presented in Figure~\ref{fig:ChargePump_schematic}.

\subsubsection{Calibration Loop Performance}
\begin{figure}[H]
    \centering
    \includegraphics[width=0.5\textwidth]{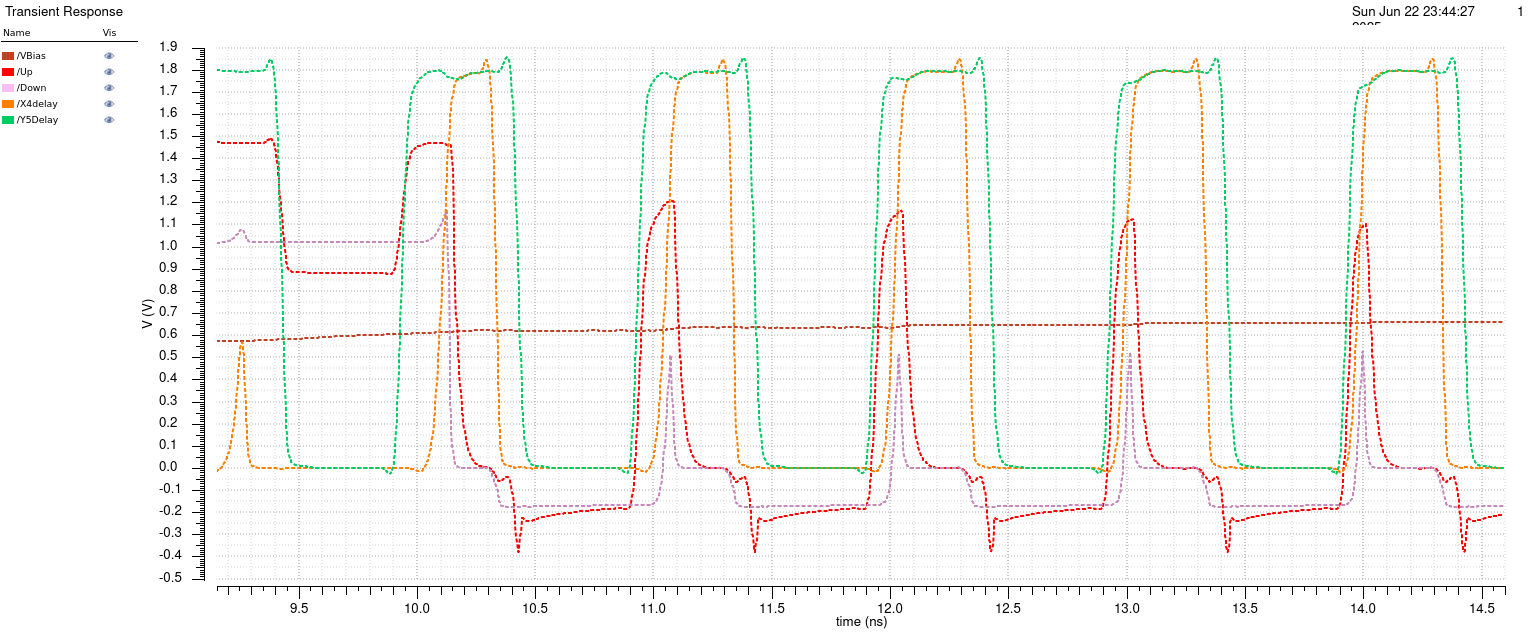}
    \caption{Transient response of the calibration loop at the start, from 9.5 ns to 14.5 ns. UP signal is larger than DN, causing \(V_{bias}\) to increase and delay mismatch to reduce.}
    \label{fig:Calibration_start_zoom}
\end{figure}

Transient simulations validated the effectiveness of the calibration loop. Initially, the UP signal dominates, causing the bias voltage (\(V_{bias}\)) to increase, thus adjusting the X delay line toward alignment, as shown in Figure~\ref{fig:Calibration_start_zoom}. 

\begin{figure}[H]
    \centering
    \includegraphics[width=0.5\textwidth]{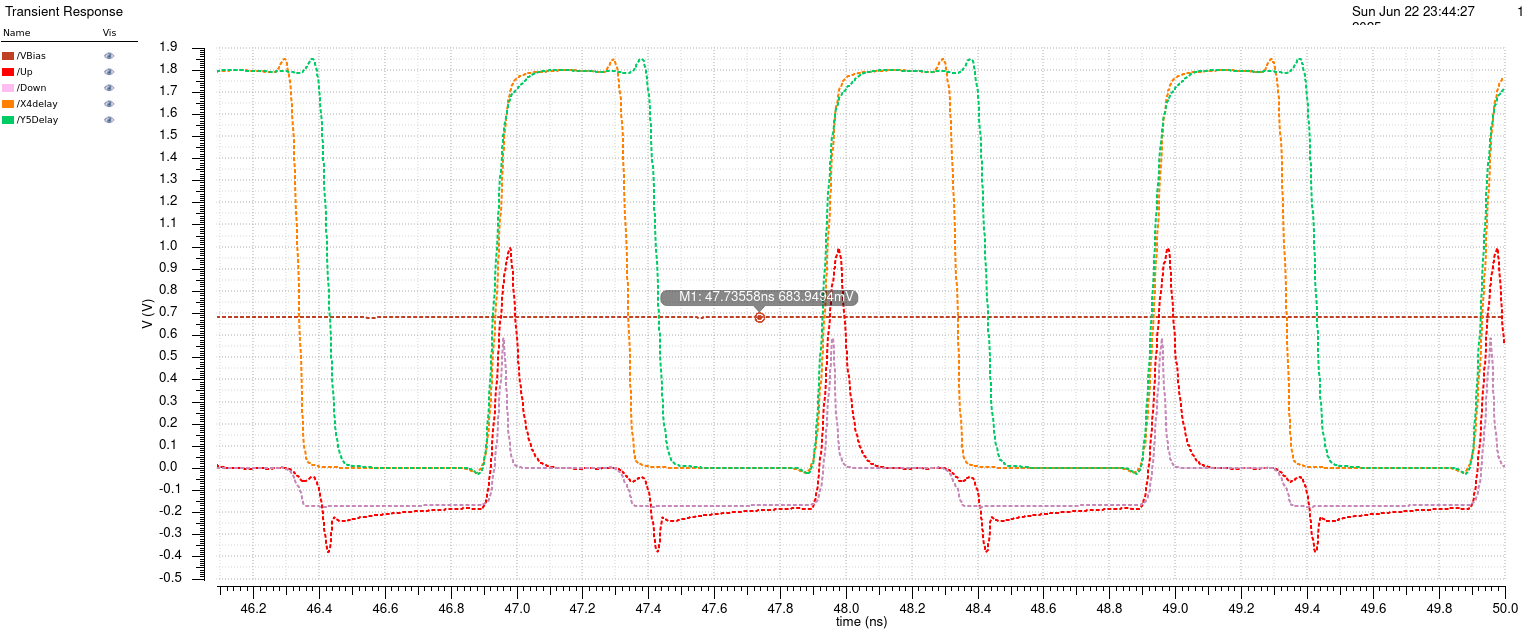}
    \caption{Transient response of the calibration loop at steady-state (approx. 50 ns). \(V_{bias}\) is stable at 683.9 mV, showing aligned start and stop signals.}
    \label{fig:Calibration_end_zoom}
\end{figure}

After sufficient calibration time, the loop reaches steady-state operation. As Figure~\ref{fig:Calibration_end_zoom} shows, \(V_{bias}\) settles to a stable level at 683.9 mV, indicating precise delay alignment between the start (X-line) and stop (Y-line) signals.

\begin{figure}[H]
    \centering
    \includegraphics[width=0.5\textwidth]{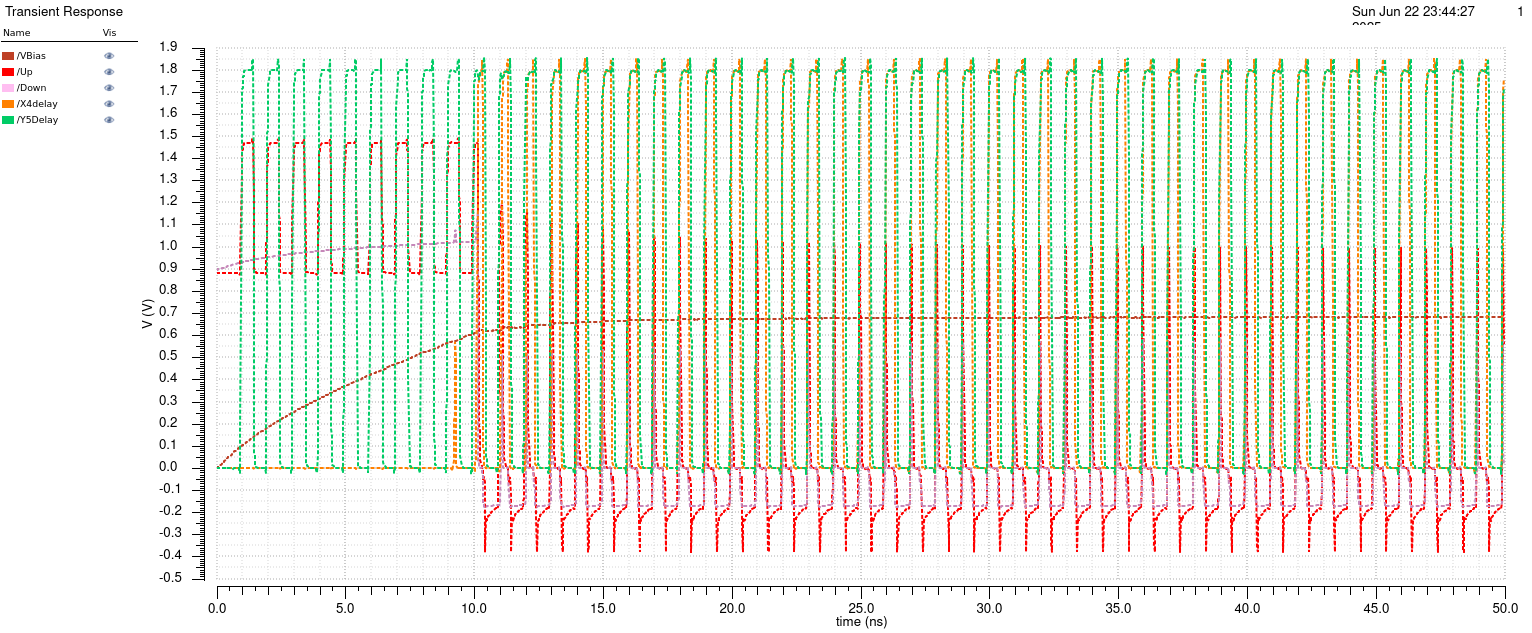}
    \caption{Overall transient response of the calibration loop, demonstrating settling behavior of \(V_{bias}\) from 0 to 50 ns.}
    \label{fig:Calibration_full_response}
\end{figure}

The overall transient behavior, from the initial mismatch to final alignment, is depicted in Figure~\ref{fig:Calibration_full_response}.

These results confirm that the designed calibration loop effectively maintains the required timing alignment, significantly improving the accuracy and robustness of the 2D Vernier TDC.
\section{Simulation and Results}
\label{Simulation_and_Results}
The simulation was performed with a time step of 5e-13, a transient duration of 5e-9s, and a supply voltage of 1.8 V. Table \ref{tab:pvt_results} summarizes the key results across different process-voltage-temperature (PVT) corners. The design consistently meets the specified requirements under all tested conditions.

All area estimations in this design are based on implementation in a \(180\,\mathrm{nm}\) CMOS technology. Transistor area is expressed in \(\mathrm{nm}^2\), as it is the product of total gate width and gate length—both measured in nanometers.

The total gate widths and number of transistors are calculated as follows:

\begin{itemize}
    \item \textbf{SR Latches:} 45 units  
    \begin{itemize}
        \item Width per latch: \(4840\,\mathrm{nm}\)  
        \item Transistors per latch: 12  
        \item Total width: \(45 \times 4840 = 217800\,\mathrm{nm}\)  
        \item Total transistors: \(45 \times 12 = 540\)
    \end{itemize}

    \item \textbf{Buffers:} 16 units  
    \begin{itemize}
        \item Width per buffer: \(18700\,\mathrm{nm}\)  
        \item Transistors per buffer: 5  
        \item Total width: \(16 \times 18700 = 299200\,\mathrm{nm}\)  
        \item Total transistors: \(16 \times 5 = 80\)
    \end{itemize}

    \item \textbf{PFD:} 1 unit  
    \begin{itemize}
        \item Total width: \(9900\,\mathrm{nm}\)  
        \item Total transistors: 8
    \end{itemize}

    \item \textbf{Charge Pump:} 1 unit  
    \begin{itemize}
        \item Total width: \(1980\,\mathrm{nm}\)  
        \item Total transistors: 6
    \end{itemize}

    \item \textbf{Inverter:} 1 unit  
    \begin{itemize}
        \item Total width: \(660\,\mathrm{nm}\)  
        \item Total transistors: 2
    \end{itemize}
\end{itemize}

\vspace{1em}
Summing all gate widths:
\[
217800 + 299200 + 9900 + 1980 + 660 = 529540\,\mathrm{nm}
\]

Summing all transistors:
\[
540 + 80 + 8 + 6 + 2 = 636
\]

The total area is computed by multiplying the total width with the gate length \(180\,\mathrm{nm}\):
\[
529540 \times 180 = 95317200\,\mathrm{nm}^2 = 95.3172\,\mu m^2
\]

\begin{table}[H]
\centering
\caption{Performance Metrics Across PVT Corners}
\label{tab:pvt_results}
\resizebox{0.5\textwidth}{!}{%
\begin{tabular}{|l|c|c|c|c|c|c|c|}
\hline
\textbf{Corner} & \textbf{FOM} & \textbf{ENOB} & \textbf{Energy} & \textbf{DNL} & \textbf{INL} & \textbf{Resolution} & \textbf{Offset} \\
\hline
tt & 0.12p & 4.95 & 4.05pJ & 0.23 & 0.20 & 16.9ps & -2.9ps \\
ss & 0.12p & 4.95 & 3.68pJ & 0.16 & 0.20 & 20.0ps & -4.27ps \\
ff & 0.15p & 4.94 & 4.60pJ & 0.19 & 0.21 & 14.9ps & -2.98ps \\
sf & 0.132p & 4.94 & 4.04pJ & 0.24 & 0.196 & 17.3ps & -3.77ps \\
fs & 0.134p & 4.93 & 4.08pJ & 0.21 & 0.22 & 17.7ps & -3.3ps \\
\hline
\end{tabular}
}
\end{table}

\begin{figure}[H]
    \centering    \includegraphics[width=1\linewidth]{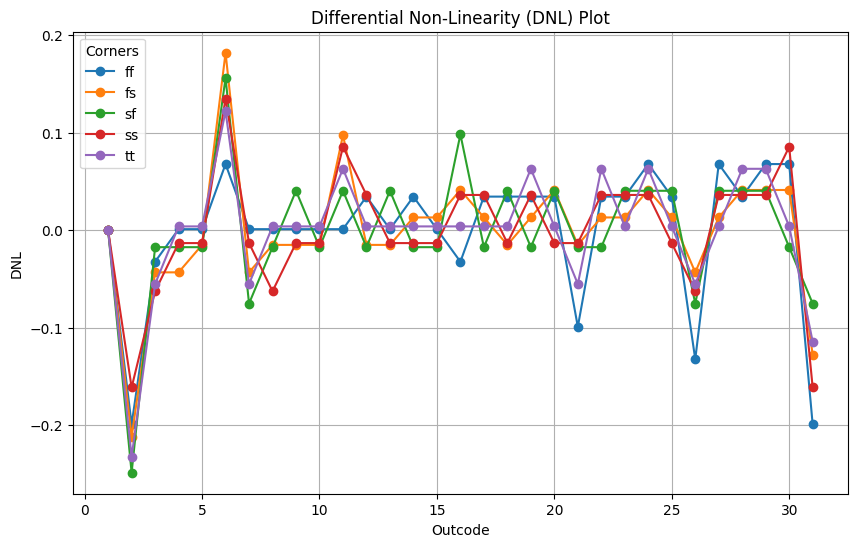}
    \caption{DNL across different corners }
    \label{fig:dnlplot}
\end{figure}

\begin{figure}[H]
    \centering
    \includegraphics[width=1\linewidth]{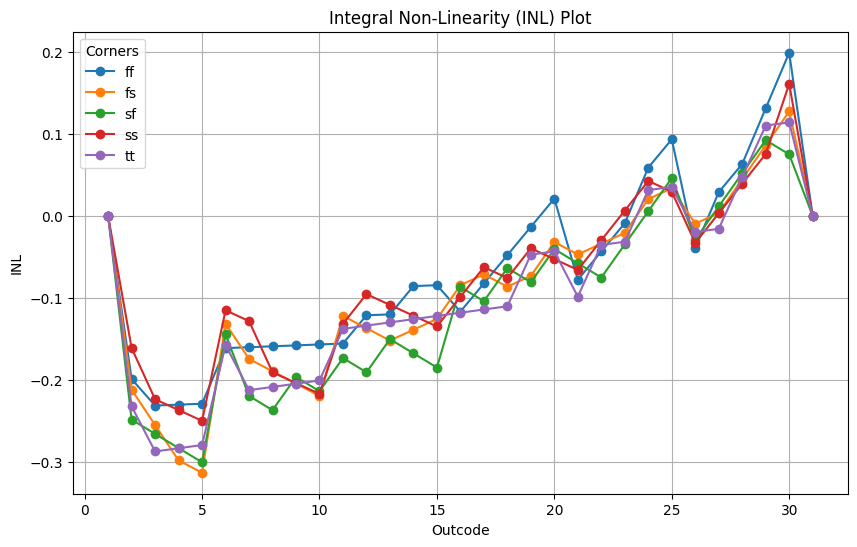}
    \caption{INL across different corners}
    \label{fig:INL}
\end{figure}

Figure \ref{fig:dnlplot} illustrates the Differential Non-Linearity (DNL) across various corners. Notice that each transition between diagonals in the Vernier plane corresponds to a peak in the DNL plot. The initial peak is likely caused by the Start signal’s rising edge after the first buffer, which is significantly steeper compared to subsequent stages. Similarly, Figure \ref{fig:INL} presents the Integral Non-Linearity (INL) plot, where peaks are also observed at each diagonal transition.
\begin{figure}[H]
    \centering    \includegraphics[width=1\linewidth]{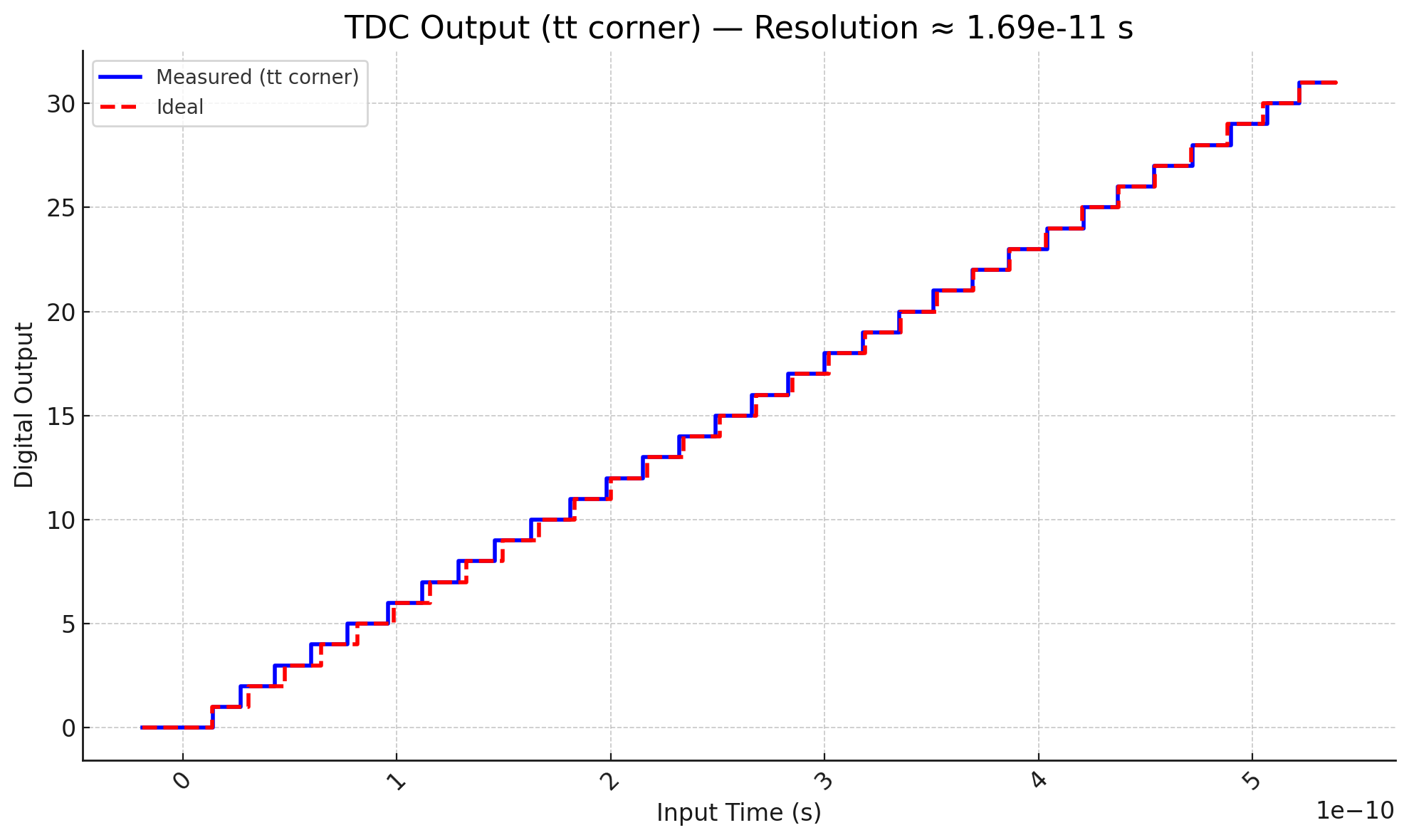}
    \caption{Transfer curve of the TDC for tt corner}
    \label{fig:tfcurvett}
\end{figure}
Figure \ref{fig:tfcurvett} shows the transfer curve of the TDC at the typical-typical (tt) process corner. The plot compares the measured digital output (blue stair-step curve) with the ideal linear response (red dashed line) over a range of input time delays.

The transfer curve illustrates the quantization behavior of the TDC, where the output increases in discrete steps corresponding to the TDC's resolution. In this case, the resolution is approximately 16.9 ps. The close alignment between the measured and ideal curves suggests good linearity and minimal non-linearity in the TDC architecture.

\section{Discussion}
The resolution of a 2D Vernier TDC is defined by the delay mismatch between delay lines X and Y. In the implemented design, the delay per stage in line Y is denoted as \( \tau \), and line X is intentionally made slower with a stage delay of \( \tau_X = \frac{5}{4} \tau \). As a result, the achieved resolution is:

\[
\Delta = \tau_X - \tau_Y = \frac{1}{4} \tau
\]

This configuration offers a good balance between resolution, power, and area.

\subsection{Improving Resolution}
To achieve finer resolution, one can increase the number of stages in lines X and Y while keeping their total delays nearly equal. If line Y has \( N_Y \) stages and line X has \( N_X \) stages, then assuming a common delay per Y-stage of \( \tau \), the delay per X-stage becomes \( \tau_X = \frac{N_Y}{N_X} \tau \). The theoretical resolution then becomes:

\[
\Delta = \tau_X - \tau_Y = \left( \frac{N_Y - N_X}{N_X} \right) \tau
\]

For example, choosing \( N_Y = 31 \) and \( N_X = 30 \) results in:

\[
\Delta = \left( \frac{1}{30} \right) \tau
\]

This is a very small fraction of \( \tau \), much smaller than the earlier case with \( \Delta = \frac{1}{4} \tau \).

\subsection{Trade-Offs}
While this strategy enables extremely fine resolution, it comes with trade-offs:

\begin{itemize}
    \item \textbf{Area Overhead:} More delay stages increase the number of SR latches quadratically. For example, \( N_X = 30 \) and \( N_Y = 31 \) requires \( 930 \) latches and \( 61 \) delay buffers.
    \item \textbf{Higher Power Consumption:} Larger arrays and longer signal paths result in higher dynamic and leakage power.
    \item \textbf{Complex Calibration:} Precise tuning is harder when the difference between stage delays becomes extremely small.
\end{itemize}

This highlights a key design trade-off: extremely fine resolution is achievable, but it must be weighed against area, power and circuit complexity.

\section{Conclusion}
The implemented design fulfills all specified performance requirements, with the calibration loop demonstrating effective operation in maintaining accurate timing relationships. The TDC exhibits robust performance across various process, voltage, and temperature (PVT) corners, consistently operating within the defined specifications. Linearity metrics show that both differential nonlinearity (DNL) and integral nonlinearity (INL) remain below 0.5 least significant bits (LSB), confirming high linearity accuracy. It is important to highlight that comparisons with the work of \cite{2dvernierplane} should be interpreted with caution, as their design utilizes a 7-bit resolution and includes a complete layout implementation, whereas the present study is limited to schematic-level simulations with a 5-bit resolution, rendering a direct performance comparison not entirely equivalent. But by following the design procedure, the implemented design meets the requirements.




\appendices
\section{Additional Figures}
\end{document}